\documentclass[english,letterpaper,aps,prl,twocolumn,showpacs]{revtex4}
\usepackage{lmodern}

\usepackage[T1]{fontenc}
\usepackage[latin1]{inputenc}
\usepackage{graphicx}
\usepackage{amssymb}

\makeatletter

\usepackage{lmodern}

\makeatletter

\usepackage{lmodern}

\makeatletter

\usepackage{lmodern}

\makeatletter

\usepackage{lmodern}

\makeatletter

\makeatletter

\usepackage{epsfig}

\usepackage{dcolumn}

\usepackage{bm}

\usepackage{bbm}

\usepackage{wasysym}

\usepackage{marvosym}

\newlength{\textwidthm}

\makeatother

\usepackage{babel}
\makeatother

\begin{document}

\title{The electron many-body problem in graphene }

\author{Bruno Uchoa, James P. Reed, Yu Gan, Young Il Joe, Eduardo Fradkin}
\affiliation{Department of Physics and Frederick Seitz Materials Research Laboratory, University of Illinois, Urbana, IL, 61801, USA}

\author{Diego Casa}
\affiliation{Advanced Photon Source, Argonne National Laboratory, Argonne, IL, 60439, USA}

\author{Peter Abbamonte}
\affiliation{Department of Physics and Frederick Seitz Materials Research Laboratory, University of Illinois, Urbana, IL, 61801, USA}

\date{\today}

\begin{abstract}
We give a brief summary of the current status of the electron many-body
problem in graphene. We claim that graphene has intrinsic dielectric
properties which should dress the interactions among the quasiparticles,
and may explain why the observation of electron-electron renormalization
effects has been so elusive in the recent experiments. We argue that the 
strength of Coulomb interactions in graphene may be characterized by an 
effective fine structure constant given by $\alpha^{\star}(\mathbf{k},\omega)\equiv2.2/\epsilon(\mathbf{k},\omega)$,
where $\epsilon(\mathbf{k},\omega)$ is the dynamical dielectric function.
At long wavelengths, $\alpha^{\star}(\mathbf{k},\omega)$ appears
to have its smallest value in the static regime, where $\alpha^{\star}(\mathbf{k}\to0,0)\approx1/7$
according to recent inelastic x-ray measurements, and the largest
value in the optical limit, where $\alpha^{\star}(0,\omega)\approx2.6$.
We conclude that the strength of Coulomb interactions in graphene is not universal, 
but depends highly on the scale of the phenomenon of interest. We propose a 
prescription in order to reconcile different experiments.
\end{abstract}


\maketitle

\section{Introduction }

Graphene is a single atomic layer of graphite, that has been isolated
only a few years ago\cite{Novoselov}, and whose elementary quasiparticle
excitations behave as massless Dirac fermions propagating in two spatial
dimensions\cite{Antonio}. An important and to the present moment
unsolved problem is determining to what extent electron-electron interactions
are important in this material. In isolated free standing samples,
a trivial estimate for the strength of electron-electron interactions
in graphene can be achieved by computing ratio of the Coulomb
energy, $E_{C}=e^{2}n^{1/2}$, where $e$ is the electron charge,
and $n$ is the electronic density, to the Kinetic energy, $E_{K}=\hbar vn^{1/2}$,
where $\hbar v=6 \, \mbox{eV\AA}$ is the velocity of the particles. Because
the dispersion of the electrons is linear, in two dimensions
those two energies scale in the same way with the density, and
their ratio is a dimensionless constant known as the fine structure
constant\cite{Kotov}, \begin{equation}
\alpha=\frac{e^{2}}{\hbar v}\approx2.2\,,\label{eq:alpha}\end{equation}
which is 300 times larger than the usual fine structure constant in
quantum electrodynamics, $\alpha_{QED}=e^{2}/(\hbar c)=1/137$, with
$c$ the speed of light. 

Although graphene is not as strongly interacting
as other materials such as cuprates, where the strength of interactions
is measured by the ratio $U/t\sim10$, where $U$ is the local Coulomb
repulsion and $t$ is the hopping energy, by this argument freestanding graphene 
should not be a weakly interacting system either. The validity of standard perturbation
theory requires $\alpha<1$, which is not applicable in suspended
samples. A  coupling constant of $\approx2.2$ is large enough to result in a complete breakdown of perturbation theory. 

Why graphene fails to exhibit dramatic correlation effects - even in two 
dimensions - is one of its most challenging puzzles.  Exacerbating the problem, 
a variety of experimental results that are sensitive to interaction effects 
{\it appear} to give contradictory results, and further appear to contradict 
what is expected from theory. Since the Fermi surface of neutral graphene (at half
filling) is formed by points at the edges of the Brillouin zone, the
Dirac points, metallic screening is not expected to influence its
electronic properties due to the vanishing density of states. At the
same time, electron-electron interactions are generally expected to
give rise to a logarithmic renormalization of the the Fermi velocity
at the Dirac points\cite{Gonzalez}, \begin{equation}
v(q)=v\left[1+\frac{\alpha}{4}\ln\!\left(\frac{\Lambda}{q}\right)\right],\label{eq:v(q)}\end{equation}
where $\Lambda$ is the bandwidth, which plays a role of an ultraviolet
cut-off, and $q$ is the momentum measured away from the Dirac point.
The structure of perturbation theory in graphene is such that this
logarithmic divergence cannot be resummed in higher order expansion
in $\alpha$\cite{Gonzalez,Mishchenko}, in contrast to the behavior
of Dirac fermions in one dimension, the so called Luttinger liquids\cite{Giamarchi}.
In addition, the electron charge $e$ does not renormalize, so the
coupling constant $\alpha$ is expected to be logarithmically renormalized
to zero near the Dirac points, i.e. the interaction is marginally
irrelevant. Although the damping of the quasiparticles is expected
to be $\tau(\omega_{0})\propto\alpha^{2}|\omega_{0}|$ to leading
order in $\alpha$, where $|\omega_{0}|=\hbar vq$ is the energy of
the quasiparticles, because of the logarithmic renormalization of
$v$ and $\alpha$ the ratio $\tau(\omega_{0})/|\omega_{0}|\ll1$
is small in the limit $|\omega_{0}|\to0$, so the quasiparticles are
still well defined (with logarithmic accuracy) near the Dirac points.
In general, the renormalization of all physical quantities, such as
the compressibility, susceptibilities, etc. can be derived directly
from their scaling dependence with the renormalized velocity\cite{Barlas,Hwang,Sheehy}.

To be more concrete, one should expect for instance the single-particle
spectrum of graphene, to be logarithmically renormalized near the
Dirac points. In supported samples, the fine structure constant is
dressed by dielectric screening effects from the substrate, $\alpha=e^{2}/(\hbar v\epsilon_{0})$,
with $\epsilon_{0}$ the dielectric constant. Angle resolved photoemission
spectroscopy (ARPES) experiments do not show any evidence of a logarithmic
renormalization in the spectrum\cite{Bostwick}, even though in graphene
on a moderate dielectric substrate such as SiO$_{2}$, where $\alpha\sim1$,
a visible renormalization should be expected in the energy range of
eV away from the Dirac point. 

Another interesting observable that can reveal renormalization effects
in graphene is the the electronic compressibility, $\kappa=(\partial^{2}E/\partial n^{2})^{-1}$,
with $E$ the total energy, which measures the strength of interactions
in an electron gas. This measurement has been recently carried out
by a single electron transistor (SET) experiment\cite{Martin}, which
revealed large puddles of change in graphene at densities as low as
$10^{11}\mbox{cm}^{-2}$, which corresponds to a Fermi level $\sim$50
meV away from the Dirac point. For non-interacting Dirac fermions,
$\kappa^{-1}=v\sqrt{\pi/4n}$. Many body effects were expected to
give an additional logarithmic correction with the density due to
the velocity renormalization\cite{Barlas}. The deviation from the
non-interacting result, nevertheless, was not observed. Although those
experiments were not carried out on suspended samples, the modest dielectric
screening from the SiO$_{2}$ substrate is not sufficient to explain
the absence of verifiable renormalization effects. To make things
more intriguing, recent ARPES measurements on doped graphene samples
observed a non-trivial splitting of the Dirac cones\cite{Bostwick2},
which were attributed to the formation of plasmarons, a composite
particle formed by an electron and a plasmon, the collective charge
excitation of the Fermi sea. 

Many-body effects have also been observed in the presence of strong
magnetic fields\cite{Checkelsky}, which quench the kinetic energy
into discrete Landau levels. The most convincing evidence of interactions
is based in the observation of $\nu=1/3$ plateau in the fractional
quantum hall effect in suspended samples\cite{Bolotin,Du,Ghahari}
and in samples supported on boron nitride\cite{Dean}. At zero magnetic
fields, on the other hand, there is no experimental evidence so far
of the excitonic gap\cite{Khveshchenko}, which has been predicted
to open up in graphene when $\alpha>1.1$\cite{Drut}. In all current
transport and spectroscopy experiments, graphene seems to behave as
a semi-metal\cite{Li,Du2}. 

From the point of view of Coulomb scatterers, experiments involving
the adsorption of K adatoms in graphene have reported a significant
change in the electronic mobility of the samples\cite{Chen}, what
was interpreted as an indication that the transport in suspended samples
was severely influenced by Coulomb scattering from charge impurities.
This interpretation is nevertheless at odds with another experiment\cite{Ponomarenko}
in which the dielectric constant of the SiO$_{2}$ substrate was enhanced
up to two orders of magnitude by embedding the substrate in highly
dielectric fluids. The variation of the mobility in this experiment
was no more than 30\%, indicating that Coulomb impurities do not influence
the electronic properties of graphene. 

As we argue below, these apparent inconsistencies among different
experiments and the difficulty of observing electronic many body effects
in the various physical observables may be due to the intrinsic dielectric
screening properties of graphene itself. In suspended samples, where
$\alpha=2.2$, interactions among the electrons can be quite strong
and lead to dynamical screening of the quasiparticles, which can be
much more weakly interacting than previously believed.

\section{The polarization of the vacuum}

As the electrons interact, virtual processes that excite electrons
from the filled valence band up to the empty conduction band spontaneously
create particle-hole pairs which eventually recombine and decay back
into the ground state, ie. the vacuum. The process of spontaneous
creation and annihilation of particles and holes polarizes the charge
of the system. This polarization is expected to dress the quasiparticles
and give rise to screening. Due to the absence of a Fermi surface
(say at half-filling), the screening is dielectric in nature, in the
sense that the Coulomb interaction remains long ranged but parametrically
weaker. For a review about electron-electron interactions and the
charge polarization of graphene, see Ref.\cite{Kotov}. 

\begin{figure}
\begin{centering}
\includegraphics[scale=0.34]{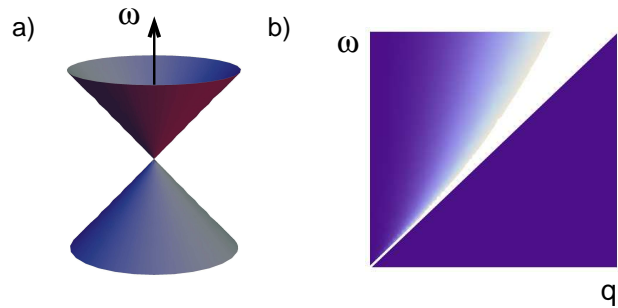}
\par\end{centering}

\begin{centering}
\caption{{\small a) Energy spectrum of the electrons around the Dirac point,
$\omega=\hbar vq$. b) Density map of the imaginary part of the polarization
for non-interacting Dirac fermions, $\mbox{Im}\Pi^{(1)}(q,\omega)$.
Dark blue color: $\mbox{Im}\Pi^{(1)}(q,\omega)=0$; white region:
optical absorption edge at the border of the particle-hole continuum
($\omega>\hbar vq$).}}

\par\end{centering}
\end{figure}

Although graphene is two dimensional, the fact that it is a semi-metal
rather than gapped allows dielectric screening to emerge at long wavelengths.
This is easily understood if one computes the charge polarization
function for non-interacting Dirac fermions in $2+1$ dimensions,
\begin{equation}
\Pi^{(1)}(q,\omega)=-\frac{1}{4}\frac{q^{2}}{\sqrt{(\hbar vq)^{2}-\omega^{2}}}\,.\label{eq:Pi}\end{equation}
This polarization function is strictly real for $\omega<\hbar vq$
and becomes purely imaginary for $\omega>\hbar vq$, in the region
of the particle-hole continuum where virtual excitation processes
from the lower to the upper band are allowed (see Fig. 1).  In the static regime
$(\omega=0$), the dielectric function of graphene in random phase
approximation (RPA) is \begin{equation}
\epsilon(q,0)=1-V(q)\Pi^{(1)}(q,0)=1+\pi\alpha/2\approx4.45\label{eq:epsilon}\end{equation}
for $\alpha=2.2$\cite{note1}, where \begin{equation}
V(q)=\frac{2\pi e^{2}}{q}\label{eq:Vq}\end{equation}
is the Fourier transform of the Coulomb interaction $V(r)=e^{2}/r$
in 2D. For gapped graphene, as in any semiconductor, there is a crossover
in the behavior of the static polarization function at long wavelengths,
$\hbar vq\ll\Delta$, $\Pi^{(1)}(q,0)=-q^{2}/\Delta$, where $\Delta$
is the energy gap\cite{Kotov2}. In two spatial dimensions, where
the Coulomb interaction $V(q)\propto1/q$, one recovers the dielectric
constant of the vacuum $\epsilon(q\ll\Delta/\hbar v,0)=1$, in contrast
with 3D insulators ($V(q)\propto1/q^{2})$, where the dielectric function saturates to a constant for $q\ll \Delta/\hbar v$. 

In the gapless case, if $\alpha=2.2$, there is no reason, a priori,
to trust in the leading correction of perturbation theory in the physical
observables, and one should seek to include diagrams of all orders
in $\alpha$. If one goes on to include the next correction to the
dielectric constant, which is enclosed in the vertex correction of
the bubble, $V(q)\Pi_{\mathrm{vertex}}^{(2)}(q,0)\approx-0.53\alpha^{2}.$
For $\alpha=2.2$, this term is almost of the same order of the RPA
correction, in which case\cite{Kotov3}\begin{equation}
\epsilon(q,0)=1+\pi\alpha/2+0.53\alpha^{2}\approx7+O(\alpha^{3})\,.\label{eq:epsilon3}\end{equation}
In the opposite regime, where $\omega\gg\hbar vq$, one recovers the
gapped situation, since the polarization bubble in leading order is
\begin{equation}
\Pi^{(1)}(q\to0,\omega)=-\frac{1}{4}\frac{q^{2}}{i\omega}\,,\label{eq:pi1}\end{equation}
and hence $\epsilon(q\to0,\omega)\to1$. Therefore, the screened Coulomb
interaction, and as a consequence the dressed fine structure constant
of graphene, \begin{equation}
\alpha^{\star}(q,\omega)\equiv\frac{e^{2}}{\hbar v}\frac{1}{\epsilon(q,\omega)}\,,\label{eq:alphastar}\end{equation}
has two distinct limits: the static regime, where $\alpha^{\star}(q,0)=\alpha_{G}<2.2$,
and the dynamic limit, where screening is ineffective and $\alpha^{\star}(0,\omega)=2.2$
recovers the bare coupling constant of freestanding samples. In the
on-shell intermediate case, where $\omega\approx\hbar vq$, the leading
vertex correction to the polarization bubble has a logarithmic divergence
near the on-shell region\cite{Gangadharaiah}, \begin{equation}
\Pi_{\mathrm{vertex}}^{(2)}(q,\omega)\propto \frac{q^{2}}{\omega-\hbar qv}\,\ln\!\left(\frac{\hbar vq}{|\omega-\hbar vq|}\right).\label{eq:Piv_q,w}\end{equation}
Resummation of this divergence in the ladder channels to all orders
in $\alpha$ has been proposed to give rise to a zero of the dielectric
function corresponding to an excitonic bound state that lives in the
optical gap of the particle-hole continuum\cite{Gangadharaiah}. In any case, the RPA approximation
seems to underestimate screening in the low energy sector $\omega\ll\hbar vq$
and does not account for excitonic effects at the edge of the particle-hole
continuum.

From the point of view of Coulomb impurities, the amount of charge
induced by the polarization of the vacuum around a test charge $Q$
can be computed directly from the dielectric function assuming linear
response,\begin{equation}
Q_{\mathrm{induced}}(q)=-Q\left(1-\frac{1}{\epsilon(q,0)}\right)\,,\label{eq:Qind}\end{equation}
where $\epsilon(q,\omega)=1-V(q)\Pi(q,\omega)$ is the dielectric
function. Since for Dirac fermions $\epsilon(q,0)$ is a constant,
the real space distribution of the induced charge is a delta function
centered at the impurity. For $\alpha=2.2$, the overall induced charge
	is $Q_{\mathrm{induced}}^{RPA}\approx-0.77Q$ in RPA, and $Q_{\mathrm{induced}}\approx-0.86Q+O(\alpha^{3})$
when the first vertex correction in the polarization is included. 

\begin{figure}
\begin{centering}
\includegraphics[scale=0.38]{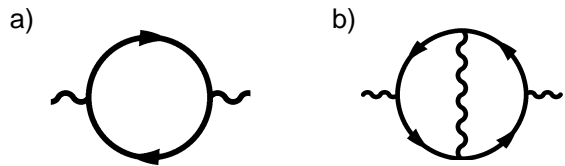}
\par\end{centering}

\begin{centering}
\caption{{\small Diagrammatic representation of a) the leading term in the
polarization function, $\Pi^{(1)}$ and b) the lowest order vertex
correction.}}

\par\end{centering}
\end{figure}

\section{Experimental measurement of the polarization}

One experiment that can provide direct information about the charge
polarization properties of a system is the measurement of the response
function with inelastic x-ray scattering. The response function is
the imaginary part of the charge susceptibility, $\chi(q,\omega)$,
which is defined in terms of the full polarization function, $\Pi(q,\omega)$,
\begin{equation}
\chi(q,\omega)=\frac{\Pi(q,\omega)}{\epsilon(q,\omega)}\,.\label{eq:chi}\end{equation}

One major difficulty with this experiment for free standing graphene
samples is that x-ray experiments cannot readily be performed on 
a one atom thick material.
This difficulty can be overcome by realizing that although the susceptibility
of the multilayer system can be quite different from the single layer
case, they have very similar polarization functions at momentum and
energy scales that are much larger than the electronic hopping energy
between the different layers. The argument is the following: let us
suppose for the moment that the interlayer hopping energy, $t_{\perp}$,
is zero. The Coulomb interaction is long ranged and couples all the
layers, \begin{equation}
V_{3D}(k)=\frac{4\pi e^{2}}{|\mathbf{k}|^{2}+k_{z}^{2}}\,,\label{eq:V3D}\end{equation}
with $\mathbf{k}=(k_{x},k_{y})$ the in-plane momentum. In the diagram
in Fig. 2b for the vextex, it is clear that since the fermionic lines
have no $k_{z}$ dependence (since the fermions do not disperse along
that direction when $t_{\perp}=0$), the integration over $k_{z}$
in the internal Coulomb line gives $\int_{-\infty}^{\infty}\frac{\mbox{d}k_{z}}{2\pi}V_{3D}(k)=2\pi e^{2}/|\mathbf{k}|$,
which is the Coulomb interaction of electrons in the single layer.
This argument can be extended to all orders in perturbation theory
in $\alpha$. Hence, the polarization functions of the freestanding
single layer and of the multilayer systems should be identical provided
they share the same single particle energy spectrum\cite{note2}. 

Since the external momentum and frequency enter as an infrared cut-off
of the momentum integrals in the polarization, provided that $t_{\perp}$
is much smaller than this energy scale, say $t_{\perp}\ll\mbox{max}(\hbar vq,\omega)$,
$t_{\perp}$ can only give subleading corrections to the polarization
but cannot affect the leading term. The similarity between the polarization
function of graphene and the multi-layer case breaks down if $t_{\perp}\sim\mbox{max}(\hbar vq,\omega)$,
in which case $t_{\perp}$ becomes the infrared cut-off itself, giving
rise to a crossover. The same rationale applies to other possible
extrinsic infrared energy scales, such as a Fermi pocket, $E_{F}$,
which shifts the chemical potential away from the neutrality point,
or an external magnetic field, $B$. Both are expected to give rise
to a crossover in the behavior of the dielectric function at sufficiently
low energy scales, but do not affect the leading term of the dielectric
function provided $E_{F},\, B\ll\mbox{max}(\hbar vq,\omega)$. Hence,
in spite of the fact that the dimensionality of the Coulomb interaction
is different in the freestanding single layer and in the multilayer
case, the polarization of the vacuum is quite similar in those two
systems at energy scales where the single particle spectrum of the
two is essentially the same. 

Examples of multilayer graphene systems include graphite in the Bernal
$AB$ stacking and other variations including turbostratic graphite,
where the layers are randomly rotated. Information about the vacuum
polarization of the single freestanding layer can be also obtained
experimentally from samples with a finite number of layers.
In single crystals of graphite, $t_{\perp}\approx0.39$eV is the interlayer
hopping energy, below which the bands have hyperbolic dispersion near
the $K$ points of the Brillouin zone. In the turbostratic case, $t_{\perp}$
can be much smaller since the coherence of the hopping between different
layers is suppressed. Multilayer epitaxial grown graphene samples
could also be suitable for x-ray experiments, and reveal detailed
information about the vacuum polarization properties of isolated graphene
in the infrared. 

Recent inelastic x-ray experiments in single crystals of graphite
conducted by the authors\cite{Reed} have revealed that freestanding
graphene is a highly polarizable system. Fig. 3 summarizes the result of
those measurements.  Fig. 3a shows the absolute value of the
screened fine structure constant for a freestanding layer, which ranges
from $\alpha^{\star}(q,0)=\alpha_{G}\approx0.142\pm0.092\approx1/7$
in the static regime ($\omega\ll\hbar vq)$ and long wavelengths,
where screening is more effective, up to $\alpha^{\star}(q,\omega)\approx2.6$
in the opposite limit, $\omega\gg\hbar vq$, where screening is weak
and collective modes such as plasmons emerge due to
the diverging density of states near the Brillouin zone boundary. The solid line
$\omega=\hbar vq$ is a guide to the eye, and indicates the edge of
the particle hole continuum in the region of the Dirac cone ($q<0.5\,\mbox{\AA}^{-1}$).
In panel 3c the static charge susceptibility it is shown as a function
of the momentum. This figure indicates that the leading 
behavior of $\chi(q,0)$ is linear in $q$, which is expected for a 
system with Dirac fermions and results in a finite screening strength
at long distances. 

\begin{figure}
\begin{centering}
\includegraphics[scale=.6]{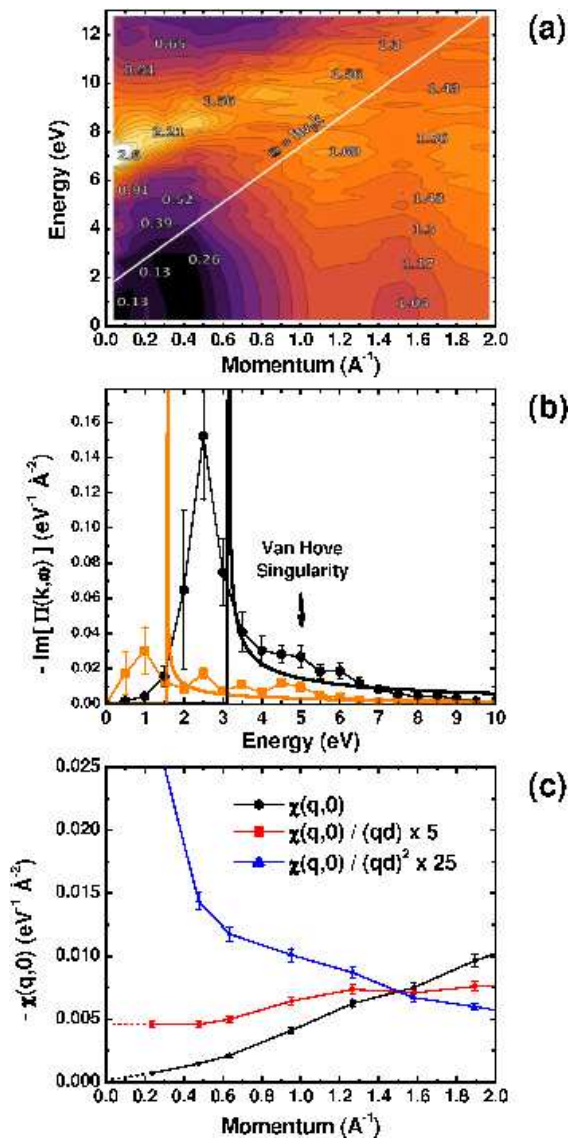}
\par\end{centering}

\begin{centering}
\caption{{\small (a) The magnitude of the effective, screened fine structure constant,
$\alpha_g^*(k,\omega)$, reproduced from ref. \cite{Reed}. (b) Imaginary part of the polarization function, 
$\Pi(q,\omega)$, measured with inelastic x-ray scattering, compared with what is expected for ideal,
Dirac Fermions (eq. 3).  (c) Asymptotic properties of the static, charge response
function, $\chi(q,0)$, showing linear, leading behavior at small $q$. }}
\par\end{centering}
\end{figure}

The imaginary part of the polarization function for the two lowest
momentum data points, $\mbox{Im}\Pi(q,\omega)$, is shown in Fig.
3b as a function of frequency and is compared with the imaginary part
of the polarization for non-interacting Dirac fermions, $\mbox{Im}\Pi^{(1)}(q,\omega)$
(solid lines), defined by Eq. (\ref{eq:Pi}). The peak in the solid
curves indicates the optical adsorption edge of the particle-hole
continuum for a fixed momentum, depicted in Fig. 1b. The data points
show a redshift of the optical adsorption edge in $\sim0.6$eV, which
is similar in magnitude with the result of a recent {\it ab initio} calculation that
explicitly accounted for interactions in the particle-hole channel\cite{Yang}.
This redshift is interpreted as an excitonic shift in spectral weight 
due to Coulomb  interaction between
electrons and holes in the conduction and valence bands respectively.
Hence, in spite of being a 2D semi-metal, graphene is polarizable to a 
degree similar to a conventional, 3D semiconductor such as Si or GaAs.

Another way to view the significant, static polarizability of graphene 
is in the screening of a test charge.
The reconstruction of the induced charge density in linear response
theory gives $Q_{\mathrm{induced}}=-(0.924\pm0.046)Q$ and the net
impurity charge is only a small fraction of the charge of the original
perturbation. In real space, the spatial distribution of the density
is a local cloud of charge with a core of approximately $R_{0}\sim10\mbox{\AA}$
of size, which quickly disappears beyond $15\AA$ away from the impurity
\cite{Reed}. Since the Dirac Hamiltonian does not have a length scale,
the corresponding charge response is purely local, as discussed below
Eq. (\ref{eq:Qind}). The high energy states by their turn 
affect the physics locally at length scales of the order the lattice
spacing and cannot not directly influence the screening properties of the system
 at long distances. 
Therefore, in half-filled graphene 
the induced charge should be completely confined
in a finite-sized region around the impurity. The implication is that 
the problem of screening a Coulomb impurity in graphene should manifest 
a complete separation of length scales.  Any crossover induced by some infrared
energy scale $E_{0}$, such as a small Fermi surface pocket, which restores
metallic screening, or a small energy gap which leads to insulating behavior,
will be manifested only beyond length scales of the order of $\lambda\sim\pi\hbar v/E_{0}$.
For a small energy scale of $E_{0}=0.1$eV measured away from the
Dirac point, the corresponding length scale is of the order of $\lambda\sim200\mbox{\AA}$.
Hence, in between these two different length scales, namely $R_{0}$
and $\lambda\gg R_{0}$, the behavior of graphene is quite universal and
screens the test charge nearly completely.

The value of the net charge found in the inelastic x-ray experiment
suggests that graphene has an intrinsic static dielectric constant
of the order of 10 {[}$\epsilon(k,0)\approx15$], which of course is 
just another way of visualizing the value given in the $\omega=0$, $q \to 0$ region of Fig. 3a. This result implies that the
transport is not significantly affected by Coulomb scatterers\cite{Ponomarenko}.
Although potassium atoms change substantially the electronic mobility
in graphene\cite{Chen}, the scattering does not seem to be driven
by the long range part of the Coulomb interaction, but rather by the
short range part of it.

\section{Prescription for physical observables}

In suspended samples, the expansion in the unscreened $\alpha=2.2$ is poorly controlled,.
The calculation of various physical observables such as the self-energy
	of the electrons and the total energy, from which one can extract
the compressibility, can be instead organized in powers of the dressed Coulomb
interaction, in the hope that the screening of the interactions among
the quasiparticles will result in a convergent expansion. 

The use of the dynamically screened Coulomb interaction in the calculation
of all physical observables may for instance reconcile apparent contradictions
between different kinds of experiments. For instance, the existence
of plasmarons in graphene, which were proposed to exist in a region
of the phase space where screening is typically weak ($\omega\gg\hbar vq)$,
is not inconsistent with the observation that Coulomb impurities can be
strongly screened out by the polarization of the vacuum, which is
related to the static response of the system ($\omega\ll\hbar vq$),
where screening is strong. The observation of the fractional quantum
Hall effect is not inconsistent either with the idea that graphene
is a strongly polarizable medium: the magnetic field in this case
plays the role of an infrared cut-off that sets an energy scale below
which the dielectric function undergoes a crossover in direction to
a {}``gapped state''. At energies comparable or smaller than this
energy scale, the vacuum is strongly reconstructed by the Landau levels,
which quench the kinetic energy and make interactions stronger. 

The lack of observable logarithmic renormalization effects in the
SET experiments on the compressibility in the single layer supported
on a SiO$_{2}$ substrate and also in ARPES experiments, can be attributed
to dynamical screening effects, which inhibit the velocity renormalization
at the energy scales currently reached by most experiments. Very recently,
the measurement of Shubnikov deHaas oscillations at very low magnetic
fields $(B\sim0.01$T) and densities as low as $10^{9}\mbox{cm}^{-2}$
have revealed the first indirect observation of the logarithmic renormalization
of the electronic spectrum in graphene\cite{Elias}. An RG calculation 
incorporating dynamically generated screening within the RPA approximation 
yielded qualitative agreement with experiment.  How
to resolve this observation with x-ray experiments, in which the breakdown of 
RPA is very clear (Fig. 3b), is not immediately obvious.  The assignment of an effective value for $\alpha$ in the Shubnikov-deHaas experiment ($\alpha_G\approx 0.6$ [29]) reflects an average over the strength of the Coulomb interactions in the entire dynamical range, and is not in principle inconsistent with the boundary values of $\alpha$ (1/7 in the static case and 2.6 in the optical regime) found in the x-ray experiment. It is perhaps suggestive that this effective value of $\alpha_G\approx 0.6$ approximately coincides with the value of the dressed fine structure constant at the edge of the particle-hole continuum [see Fig. 3 (a)], where on shell processes should dominate the renormalization of the spectrum.  Further infrared experiments are needed to verify the actual level of agreement between those two experiments.

This work was supported by the U.S. Department of Energy under grants 
DE-FG02-07ER46459 and DE-FG02-07ER46453 through the Frederick Seitz
Materials Research Laboratory, with use of the Advanced Photon Source 
supported by DEAC02-06CH11357.


\begin{thebibliography}{10}
\bibitem{note1}Of course, since $\alpha$ runs to zero logarithmically,
one should eventually expect $\epsilon(q,0)\to1$ at very long wavelengths.
\bibitem{note2}Although the argument considered a continuum of layers,
it can be easily generalized to the discrete case as well, by noticing
that the Coulomb interaction in this situation is given by $V_{3D}(k)=2\pi e^{2}dS(k)/|\mathbf{k}|$,
which is the Fourier transform of the $e^{2}/r$ interaction for an
infinite stack of layers separated by a distance $d$, and $S(k)=\mbox{sinh}(|\mathbf{k}|d)/[\mbox{cosh}(|\mathbf{k}|d)-\cos(k_{z}d)]$
is the structure factor. Integration over $k_{z}$ in the Brillouin
zone of size $2\pi/d$, namely $\int_{-\pi/d}^{\pi/d}\frac{\mbox{d}k_{z}}{2\pi}V_{3D}(k)=2\pi e^{2}/|\mathbf{k}|$,
which again is the Coulomb interaction in 2D.
\end{thebibliography}
\end{document}